\documentclass[prl,twocolumn,aps,showpacs]{revtex4}
\usepackage{dcolumn}
\input epsf
\usepackage{epsfig}

\begin{document}
\title{Direct detection constraints on superheavy dark matter}
\author{Ivone F.\ M.\ Albuquerque}\thanks{Electronic mail: IFAlbuquerque@lbl.gov}
\affiliation{Space Science Laboratory and Astronomy Department, 
        University of California, Berkeley, California \ 94720 }

\author{Laura Baudis}\thanks{Electronic mail: lbaudis@stanford.edu}
\affiliation{Department of Physics, Stanford University, 
Stanford, CA 94305.}

\date{10 Jan 2003}

\begin{abstract}
The dark matter in the Universe might be composed of superheavy particles
(mass $\vcenter
     {\hbox{$>$}\nointerlineskip\hbox{$\sim$}} 10^{10}$ GeV).
These particles can
be detected via nuclear recoils produced in elastic scatterings from nuclei.
We estimate the observable rate of strongly interacting supermassive 
particles (simpzillas) in direct dark matter search experiments.
The simpzilla energy loss in the Earth and in the experimental shields is taken into 
account. The most natural scenarios for simpzillas are ruled out based on recent 
EDELWEISS and CDMS results. 
The dark matter can be composed of superheavy particles 
only if these interact weakly with
normal matter or  if their mass is above $10^{15}$ GeV.
\end{abstract}


\maketitle

The dark matter in the Universe might be composed of supermassive particles
(mass $\vcenter
     {\hbox{$>$}\nointerlineskip\hbox{$\sim$}} 10^{10}$ GeV). 
Although the leading dark matter candidate  is a weakly interacting
massive particle (WIMP), recent models for nonthermal production of particles
in the Early Universe have broadened the dark matter mass and cross section 
parameter space.
Supermassive particles avoid the unitarity mass bound \cite{unit} by
not being thermal relics of the Big Bang. 
There are many ways to produce supermassive particles in the Early Universe. 
The most general way is gravitational production at the end of inflation
as a result of the expansion of the Universe \cite{rocky,kuzmin,lam}. 
In this scenario the average particle density is independent of the
interaction strength with normal matter. 

A reasonable assumption for the mass of such particles is the inflaton mass scale,
$\sim 10^{12}$ GeV in chaotic inflation models. These 
particles might be composed
of exotic quarks or other new particles \cite{esteban,moharaby}.
Their production allows them to be  electrically charged but 
there are arguments that they have to be neutral \cite{esteban}.

If the non-luminous matter is indeed composed of supermassive particles
its interaction strength with normal matter can 
range from weak (wimpzillas \cite{rkdark}) to strong (simpzillas \cite{ahk}). 
Although not many experiments have been built to search for 
strongly interacting dark matter, 
observations from several experiments  constrain their mass and cross section 
parameter space \cite{stark,mcguire}. Masses
up to about $10^4$ GeV are ruled out and several
allowed regions were found above this mass for a cross section range
of about $10^{-32}$ to $10^{-15}$ cm$^2$. Stronger constraints can be found in \cite{bacci}.

Here we consider the direct detection of simpzillas. 
Direct detection experiments measure the energy deposited by a  
nuclear recoil produced in the particle scaterring from a nucleus. 
Their main goal is to detect WIMPs. We show that
direct detection experiments are also able to 
probe simpzillas as a dark matter candidate.  
Comparison of our estimated simpzilla detection rate with
the latest CDMS \cite{cdms02} and EDELWEISS \cite{edelw,chardin} results, 
rules out the most natural scenarios of the simpzilla parameter space.

A simpzilla arriving at an underground detector
will have interacted many times in the Earth and, depending on its interaction
cross section with ordinary matter, will interact many times in the detector itself.
To determine if simpzillas can be directly detected,
their energy degradation and range in the Earth and the experimental shield 
has to be taken into account. 

We assume that the local dark matter is composed 
entirely of simpzillas, with a density of 0.3 GeV/cm$^3$.
The total energy deposited (Q) in a detector is given by
the nuclear recoil energy ($E_R$) times the number of simpzilla interactions
in the detector $N_I \sim n_{N} \sigma_{\chi N} l$, where $n_{N}$ is the detector
atomic number density, $\sigma_{\chi N}$ is the simpzilla nucleus cross section
and $l$ is the detector thickness. We assume that the simpzilla nucleus cross section 
is independent of the nuclear spin and relates to
the simpzilla nucleon cross section ($\sigma_{\chi p}$) by 
$\sigma_{\chi N} = \sigma_{\chi p}\,(A\,m_r / m_{rp})^2$, 
where $A$ is the nucleus atomic number, $m_r$ is the 
nucleus -- simpzilla reduced mass given by $m_\chi m_N / (m_\chi + m_N)$, 
$m_\chi$ is the simpzilla mass, $m_N$ is the nucleus mass and 
$m_{rp}$ is the nucleon -- simpzilla 
reduced mass. As $m_\chi >> m_N$, the reduced masses are simply the nucleus and the nucleon mass, respectively.

The nuclear recoil energy is given by
\begin{equation}
\label{eq:recen}
E_R = \frac{|\vec{q}\, ^2|}{2 m_N} = m_N v^2\left(1 - \cos\theta \right)
\end{equation}
where $q$ is the  momentum transferred, $v$ is the dark
matter velocity and $\theta$ is the scattering angle in the center-of-momentum
frame. For  a given material and scattering angle, the recoil energy 
depends only on the simpzilla velocity.

A simpzilla with a cross section of 10$^{-26}$cm$^2$ will interact about 10$^4$ times in a 1\,cm thick Ge detector,
while it will interact only once if the cross section is  10$^{-30}$cm$^2$. The total mean energy depositions 
will be about 450 MeV and 45 keV respectively.

We assume that the dark
matter has a Maxwellian velocity distribution ($f(v)$) with an average velocity $v_0$ 
of 220 km/s and include the motion of the 
Earth ($v_\oplus$) as described in \cite{jkg,lewin}:
\begin{equation}
\label{eq:vdis}
f(v)dv = \frac{v dv}{v_\oplus v_0 \sqrt{\pi}} \left\{ \exp\left[-\frac{(v - v_\oplus)^2}
{v_0^2}\right] - \exp \left[-\frac{(v + v_\oplus)^2}{v_0^2} \right] \right\} 
\end{equation}

The distance ($l$) traveled through the Earth to the detector is a function
of the arrival angle  between 
the simpzilla arrival direction at the Earth and the normal direction from the detector 
to the Earth's surface.
We assume simpzillas are isotropically distributed in the galactic halo. 
The average simpzilla energy at the detector is given by:
\begin{equation}
\label{eq:endeg}
E = E_0 \exp \left( -\, \frac{2 \rho N_A \sigma_{\chi N}}{m_\chi} \,l \right)
\end{equation}
where $E_0$ is the simpzilla energy at the Earth, $\rho$ is the Earth density and
$N_A$ is Avogadro's constant. We use the Earth density profile given by the
Preliminary Earth Model \cite{terra,gandhi96} and its 
composition found in \cite{comp}.

Knowing the
energy loss in the Earth we determine the maximum distance a
simpzilla can travel before being stopped. This distance is related to the
maximum arrival angle and defines an acceptance cone. 
Any simpzilla which reaches the Earth at an arrival angle larger than
this cone angle will have lost all its energy before reaching the detector. 
This cone depends on the $m_\chi$ and on $\sigma_{\chi p}$. 
The fraction of the Earth contained in the acceptance cone ($f_\oplus$) will therefore 
define the fraction of incident dark matter particles that reach the detector. 

Within an acceptance cone
the average simpzilla velocity ($v_{0d}$) and energy ($E_{0d}$)
at the detector are determined.  
The fraction of these events ($f_D$) with energy 
below a certain threshold is given by the ratio of the  lower 
velocity events over the total number of events:
\begin{equation}
\label{eq:flow}
f_D = \frac{\int^{\sqrt{2 E_{max} / m_\chi}}_{0} f(v) dv}{\int^{\sqrt{2 E_{esc} / m_\chi}}_{0} f(v) dv}
\end{equation}
where $E_{max}$ is the maximum energy which can be detected in a given detector
and $E_{esc}$ is the maximum simpzilla velocity  which we
assume to be 650 km/s, the galactic halo escape velocity.

The deposited energy spectrum $dR/dQ$ in the detector is given by:
\begin{equation}
\label{eq:drdq}
\frac{dR}{dQ} = k \left( \frac{dR_1}{dQ} - \frac{R_0}{4 E_{0d} m_N}\exp(-\frac{v_{esc}^2}{v_{0d}^2})
\right)
\end{equation}
where $k$ is a parameter in the velocity distribution \cite{lewin}
\begin{equation}
k = \left[ erf\left(\frac{v_{esc}}{v_{0d}}\right) - \frac{2 v_{esc}}{\sqrt{\pi} v_{0d}} 
\exp(-v_{esc}^2/v_{0d}^2) \right]^{-1}
\end{equation}
$dR_1/dQ$ is the differential rate including the Earth's motion \cite{lewin},
\begin{equation}
\frac{dR_1}{dQ} = \frac{R_0}{16 E_{0d} m_N} \frac{\sqrt{\pi} v_{0d}}{v_\oplus}
\left[erf\left(\frac{v_t + v_\oplus}{v_{0d}} \right) - erf\left(\frac{v_t - v_\oplus}{v_{0d}} 
\right)\right]
\end{equation}
where $v_t = \sqrt{2 E_t / m_\chi}$ and 
\begin{equation}
\label{eq:r0}
R_0 = f_\oplus f_D F^2(Q) \frac{2 N_A \rho_0}{\sqrt{\pi} A m_\chi} \sigma_{\chi N} v_{0d}.
\end{equation}
$R_0$ includes the fraction of the dark matter which is detectable. 
Besides containing the terms $f_\oplus$ and $f_D$  
which are related to strongly interacting particles, and taking the energy loss into account 
(which makes $v_{0d} < v_0$), the
differential energy spectrum given by Equation~\ref{eq:drdq} is the same as for WIMPs  
\cite{lewin}. 
For a  simpzilla -- nucleus interaction the wavelength $\lambda = \hbar / q$ 
is smaller than the nuclear radius. 
The drop in the effective cross section with increasing q 
is described by a form factor, $F^2(Q)$.  
We assume that $F^2(Q)$ is well approximated \cite{jkg,lewin} by the 
Woods-Saxon form factor \cite{engel}. 

We consider the CDMS and the EDELWEISS experiments, which employ Ge 
and, in the case of CDMS, Si detectors operated at mK temperatures 
\cite{cdms,cdms02,edelw}.
Currently, CDMS is located at a shallow depth of about 16\,mwe at the Stanford Underground 
Facility (SUF), its final destination is the Soudan mine in Minnesota, at a depth of about 
2080\,mwe.
EDELWEISS is located in the Laboratoire Souterrain de Modane at 4500\,mwe. 

A particle interacting in a cryogenic Ge or Si detector
will create phonons and electron-hole pairs. While phonons are a measure of the total 
recoil energy, only a fraction of this energy is dissipated into ionization. 
The simultaneous measurement of the phonon and ionization signals
results in an excellent discrimination efficiency against electron recoils, which are caused by
the dominant radioactive background. This discrimination is possible because 
the ratio of ionization to recoil energy is lower for nuclear recoils than for electron recoils.

We  estimate the  simpzilla elastic scattering rates at both shallow and deep sites.
We take into account the simpzilla energy loss in the experimental shields, made 
of  Cu and Pb. For CDMS, we also estimate the energy deposition in the active 
muon shield, made of plastic scintillator.
To determine the simpzilla detection rate we integrate the differential energy
spectrum given by Equation~\ref{eq:drdq}.

\begin{figure} 
\centering\leavevmode \epsfxsize=250pt \epsfbox{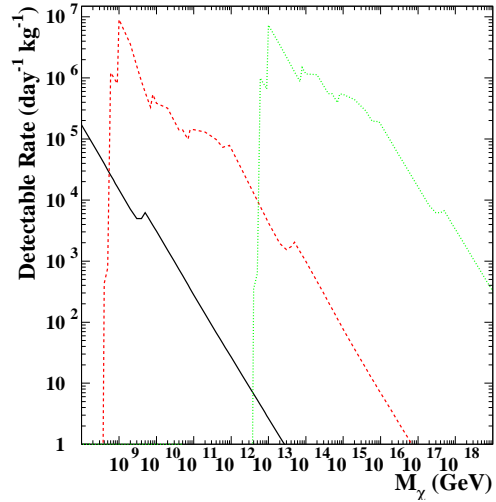}
\caption{Estimated simpzilla rate versus mass for simpzilla nucleon cross
sections of $10^{-30}$, $10^{-26}$ and $10^{-22}$ cm$^2$ (from left to right)
in the CDMS experiment at the Stanford 
Underground Facility. Kinks of each curve are due to the 
shallow depth and to the Earth density profile.}
\label{fig:det}
\end{figure}

Figure~\ref{fig:det} shows the detectable simpzilla rate versus simpzilla mass
for various cross sections, for CDMS  at SUF. 
Two effects are responsible for the kinks in these curves. First, the fact the experiment is at a 
shallow depth makes the distance
the simpzilla travels in the Earth very small for arrival angles smaller than $90^0$ and
much larger above $90^0$. Second, the simpzilla energy loss rate increases with depth, 
due to the increase in the Earth density.

The simpzilla rate at shallow sites is slightly higher than
at deeper sites. As an example, the rates of $10^{12}$ GeV simpzillas arriving
at a CDMS detector at SUF for $\sigma_{\chi p}$ of $10^{-30}$ and $10^{-26}$~cm$^2$
are 27/day/kg and $21 \times 10^4$/day/kg respectively while 4/day/kg 
and $17 \times 10^4$/day/kg respectively at the EDELWEISS site.

To obtain limits on simpzilla masses and cross sections, we compare the predicted 
simpzilla detection rates
with the background rates of the CDMS and EDELWEISS experiments in different energy 
regions. We require that the detectable rate translates into at least one
particle going through the detector in the total exposure time.
We emphasize here that our results will be conservative, for we are not attempting a 
simpzilla-specific analysis of the data.

We estimate the simpzilla rate in the 10-100\,keV recoil energy 
region for CDMS, and in the  30-100\,keV region for EDELWEISS.
$E_{max}$ in Equation~\ref{eq:flow} is set to 100 keV. We require a
minimum nuclear recoil of 1 keV per interaction, since 
the ionization efficiency in Ge decreases rapidly \cite{slrep}
below this energy.
This recoil energy range probes the region where $\sigma_{\chi p}$ 
is lower than about $10^{-28}$ cm$^2$. 
The $\sigma_{\chi p} = 10^{-30}$ cm$^2$ curve
in Figure~\ref{fig:det} is representative of the predicted rate in this 
recoil energy region.
For CDMS, we also estimate
the energy deposited in the 4.1\,cm thick plastic scintillator surrounding the detectors.
Simpzillas with a cross section below $10^{-24}$ cm$^2$, will deposit less
than 2\,MeV in this active muon shield
and thus belong to the category of muon-anticoincident events. 

\begin{figure}
\centering\leavevmode \epsfxsize=250pt \epsfbox{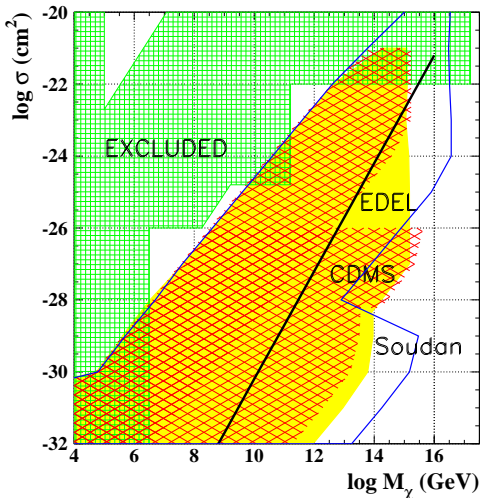}
\caption{Excluded regions at 90\% CL in
the simpzilla mass versus cross section parameter
space. The region labeled
``excluded'' was excluded in the analysis given in \cite{mcguire}; 
the regions labeled ``EDEL'' (filled) and ``CDMS'' (hatched) are excluded based 
on EDELWEISS \cite{edelw01,edelw} and CDMS results \cite{cdms02}.
The area labeled ``Soudan'' (within solid line) is the predicted region to be probed by 
CDMS at Soudan. As a comparison, the region above the thick 
straight line shows the sensitivity of future, cubic-kilometer sized neutrino
telescopes \cite{als}.}
\label{fig:simpallow}
\end{figure}

CDMS observed a total of 27 nuclear recoil events (single and multiple-scatters) 
in the 10-100\,keV region for an exposure of 15.8\,kg-day  \cite{cdms02}. 
This yields a background of 1.7 events/kg/day.  
Comparing the above number with the number of expected simpzillas in this energy region 
excludes the region labeled 'CDMS' in Figure \ref{fig:simpallow} at 90\% CL, 
for cross sections below $10^{-28}$ cm$^2$.
 
EDELWEISS has reported their result \cite{edelw} for WIMP searches based on an    
exposure of 11.7\,kg-day. It is a combined result from two measurements. The first
\cite{edelw01}, using a 320\,g 
Ge detector, had an effective exposure of 4.53\,kg-day.
 No nuclear recoils were found in the 30--200\,keV energy range. 
The second measurement 
with a new 320\,g Ge detector \cite{edelw}, had an effective exposure of 7.4\,kg-day 
and observed no nuclear recoil events from 20--100 keV. Their expected background 
from neutron scatters is about 0.03 events/kg/day  above 30\,keV \cite{edelw01}. 
The combined result corresponds to a rate inferior to 0.2~kg$^{-1}$~day$^{-1}$
at 90\% confidence level.
Comparing our estimated rate for the EDELWEISS detector with the 0.2 events/kg/day 
90\% CL limit for the 30-100\,keV recoil energy region, we exclude the region labeled 
'EDEL' in Figure \ref{fig:simpallow}, again for $\sigma_{\chi p}$ below 
$10^{-28}$ cm$^2$.
 
%%%%%%%%%%%%%%%%%%%%%%%%%%%%%%%%%%%%%%%%%%%%%%%%%%%%%%%%%%%%%%%%%%%%%%%%%%%%%%%%%%%%%%%%%%
%%%%%%%%%%%%%%%%%%%%%%%%%%%%%%%%%%%%%%%%%%%%%%%%%%%%%%%%%%%%%%%%%%%%%%%%%%%%%%%%%%%%%%%%%%
 
For CDMS, we estimate the simpzilla rates in the 3-10\,MeV ionization energy region, 
and compare it to the background in the same region shown in Figure 33 in \cite{cdms02}. 
The muon-anticoincident background in this region is very low, due to the fact that the 
background from natural radioactivity ends around 2615\,keV. 
We approximate the ionization energy to one third of the recoil energy.
The 1 keV minimum nuclear recoil energy requirement allows to probe $\sigma_{\chi p}$ of $10^{-26}$ cm$^2$
and lower. 
This comparison excludes, at 90\% CL, the $\sigma_{\chi p}$
region from about $10^{-29}$ to $10^{-26}$ cm$^2$ labeled ``CDMS'' in 
Figure~\ref{fig:simpallow}.

Simpzilla with cross sections greater than 10$^{-26}$ cm$^2$ will interact more that 
10$^4$ times in 
a 1\,cm thick Ge detector, and thus deposit an energy above 10\,MeV for a 1\,keV threshold
per interaction. In spite of their high energy, such events would 
nonetheless trigger the CDMS and EDELWEISS detectors, giving rise to
saturated pulses. 
To obtain a conservative exclusion region for these cross sections 
we consider the total background trigger rate of 0.4\,Hz \cite{cdms02} for CDMS 
and the trigger rate of  20 events/kg/day above an energy of 1.5\,MeV \cite{chardin} 
for EDELWEISS.
Since CDMS triggers on the phonon signal \cite{cdms02}, we can relax 
the threshold per interaction to 1\,eV. We assume a low-energy threshold of 5\,keV.
This excludes the region above $10^-26$ cm$^2$ in Figure~\ref{fig:simpallow}.
 
Figure~\ref{fig:simpallow} also shows the predicted 
region to be probed by CDMS at the Soudan site. We assume 
a total exposure of 230\,kg\,day and a muon--anticoincident nuclear recoil 
background of 0.02 events/kg/day in the 10-100\,keV energy region.
We also assume an overall trigger rate which is 100 times lower than the 
one at SUF.

%%%%%%%%%%%%%%%%%%%%%%%%%%%%%%%%%%%%%%%%%%%%%%%%%%%%%%%%%%%%%%%%%%%%%%%%%%%%%%%%%%%%%%%%%%
%%%%%%%%%%%%%%%%%%%%%%%%%%%%%%%%%%%%%%%%%%%%%%%%%%%%%%%%%%%%%%%%%%%%%%%%%%%%%%%%%%%%%%%%%%

We have investigated the direct detection of strongly interacting 
supermassive particles. 
We have estimated the differential and total direct  
detection rates for the CDMS and  EDELWEISS experiments. 
We find that although the energy loss as well as the depletion in the 
number of simpzillas 
reaching an underground detector are substantial, the predicted nuclear recoil rates  
are much higher than for supersymmetric WIMPs. 

Comparison of our predicted rates for CDMS at SUF
and for EDELWEISS with their most recent results
\cite{cdms02,edelw,chardin} rules out the most natural simpzilla scenarios.
The most natural scenarios are the ones for which
the simpzilla mass is comparable to the inflaton mass in chaotic inflation
models ($\sim 10^{12}$ GeV) and the simpzilla -- nucleon cross section is
comparable to the nucleon -- nucleon strong interaction cross section.
The simpzilla mass versus cross section parameter space 
which is probed by these two experiments is shown in Figure~\ref{fig:simpallow}.

The region to be tested by cubic kilometer neutrino telescopes 
such as IceCube \cite{als} is also shown. 
These telescopes can search
for secondary neutrinos produced in simpzilla annihilation in the Sun \cite{ahk,crotty}.
Although most of this region is excluded by direct detection experiments, neutrino telescopes 
would provide an independent confirmation.

In conclusion, the dark matter in the universe may be composed of superheavy 
particles only
if these particles interact weakly with normal matter or if their mass is above  $10^{15}$ GeV
or higher, depending on the simpzilla -- nucleon interaction cross section.

We thank Gabriel Chardin, Willi Chinowsky and Richard Schnee for critical comments on the manuscript.
IA was supported by NSF Grants Physics/Polar Programs 0071886 and KDI 9872979. LB was supported by the 
DOE Grant DE-FG03-90ER40569.

\end{document}